# Deciphering Mechanoluminescence: How the Nature of Mechanical Stress and Structural Dimensionality Shape Mechanisms and Responses

Alexis Duval[1,a,*], Xavier Rocquefelte[2,*], Yann Gueguen[1], Patrick Houizot[1] and Tanguy Rouxel[1,*]

**Affiliations**

[1]Univ Rennes, CNRS, IPR (Institut de Physique de Rennes) - UMR 6251, F-35000 Rennes, France

[2]Univ Rennes, CNRS, ISCR (Institut des Sciences Chimiques de Rennes) - UMR 6226, F-35000 Rennes, France

[a]Present address: Otto Schott Institute of Materials Research, Friedrich Schiller University Jena, Lessingstraße 14, 07743 Jena, Germany

[*]Author to whom correspondence should be addressed: alexis.duval@uni-jena.de; xavier.rocquefelte@univ-rennes.fr; tanguy.rouxel@univ-rennes.fr

**Abstract**

Mechanoluminescent materials exhibit a broad spectrum of controllable light-emission responses to mechanical stimuli of varying types and magnitudes. Yet progress toward high-performance systems remains constrained by an incomplete and often contradictory mechanistic understanding. Here, density functional theory (*DFT*) calculations optimized for the quantitative treatment of point defects are used to systematically investigate the interplay between stress type (hydrostatic vs. shear) and active-phase dimensionality (1D vs. 3D), using $SrAl_2O_4:Eu^{2+}, Dy^{3+}$ and $Ba_4Si_6O_{16}:Eu^{2+}, Ho^{3+}$ as representative model systems. Two distinct emission-driving mechanisms are identified: a piezoelectric contribution, and a second, apparently universal, mechanism arising from stress-induced structural reorganization at point defects sites. These results establish a design framework for mechanoluminescent materials in which crystal dimensionality, stress type and stress-sensitive point defects are deliberately matched to tune emission behavior and overall performance.

## 1. Introduction

Mechanoluminescence, *i.e.*, the emission of light in response to mechanical stimulation, has recently attracted considerable attention, with numerous proof-of-concept demonstrations highlighting its potential across remote multi-modal sensing, structural architecture monitoring, smart lighting and displays, and biomedical imaging.[1–5] Indeed, the diversity of responses (emission spectrum, intensity, lifetime) that this phenomenon offers through fine tailoring of the matrix's composition, microstructure, and crystallinity allows to meet both ubiquitous and niches demands.[6–10]

The mechanistic understanding of mechanoluminescence is closely tied to that of persistent luminescence (PersL), wherein light emission arises from the delayed release of charge carriers trapped at specific point defects within crystalline structures following prior excitation (commonly using *UV* light).[11–14] The depth of the corresponding energy levels is conventionally quantifiable by means of thermally stimulated luminescence (*TSL*) experiments, though such methodologies remain somewhat experimentally demanding and often time-consumming.[15–19]

Nevertheless, such experiments alone are insufficient for ascribing the identified energy levels to the key processes occurring at the atomic and electronic scales. In particular, the precise nature of the active point defects (whether anionic or cationic vacancies, or more complex mixed configurations) and of the charge carriers involved (electrons $e^-$ and/or holes $h^+$) often remains undetermined. In this context, the approach introduced by Freysoldt *et al.* in 2013 provides a means, through appropriate density functional theory (*DFT*) calculations, to evaluate two key properties of virtually any point defect in a solid: its formation energy (how readily the defect forms) and its trap depth (how strongly it captures and retains charge carriers).[20] Once applied to the well-known $SrAl_2O_4{:}Eu^{2+}, Dy^{3+}$ phosphor in 2023, Hoang effectively compiled and identified the point defects most likely to drive its PersL mechanism.[21]

Despite its potential, this approach has rarely been applied to elucidate the mechanism(s) underlying mechanoluminescence, which remains actively debated. In particular, the role of the piezoelectric field (and by extension the symmetry of the active crystal phase, whether centrosymmetric or not) has been identified as a key parameter governing this phenomenon,[1,22–24] notably by reducing the depth of the active energy levels in PersL and thereby enhancing the detrapping rate of the charge carriers. However, no general mechanistic picture has yet emerged. In particular, it remains unclear how centrosymmetric phases and long-range disordered systems such as glasses [25–27] are nonetheless able to exhibit mechanoluminescence.

The present study continues a series of investigations performed on crystalline $Ba_4Si_6O_{16}{:}Eu^{2+}, Ho^{3+}$ and its glass-ceramic counterparts, focusing on their photoluminescence, PersL, mechanoluminescence properties, as well as their theoretical modeling.[8,28,29] Particular emphasis is placed on quantifying the respective roles of the nature and magnitude of the applied mechanical stress – be it hydrostatic, shear, or a superposition thereof – in governing mechanoluminescence behavior and intensity during both loading and unloading. Indeed, depending on the active crystalline phase, the introduction of shear stress in addition to hydrostatic stress – under otherwise comparable loading conditions – either (i) drastically alters the mechanoluminescence behavior during unloading, with intensity increasing rather than decreasing (see $SrAl_2O_4{:}Eu^{2+}, Dy^{3+}$ in **Figure 1(a-c)**), or (ii) reduces the mechanoluminescence intensity by several orders of magnitude (see $Ba_4Si_6O_{16}{:}Eu^{2+}, Ho^{3+}$ in **Figure 1(d-f)**).[8,24,29,30]

To this end, the aforementioned method is employed to assess the sensitivity of point defects to varying mechanical stimuli. The specific case of oxygen vacancies serving as hole traps within silicate phosphors is addressed. Furthermore, a non-piezoelectric contribution – apparently universal – arising from stress-induced structural reorganization in the vicinity of point defects is identified, complementing the previously evidenced piezoelectric contribution. Importantly, the symmetry of the crystal phase (centrosymmetric or not), its structural dimensionality (1*D*, 2*D* or 3*D*), and the stress-sensitivity of point defects are shown to critically govern charge carrier relaxation dynamics and, consequently, the observed mechanoluminescence behavior and intensity.

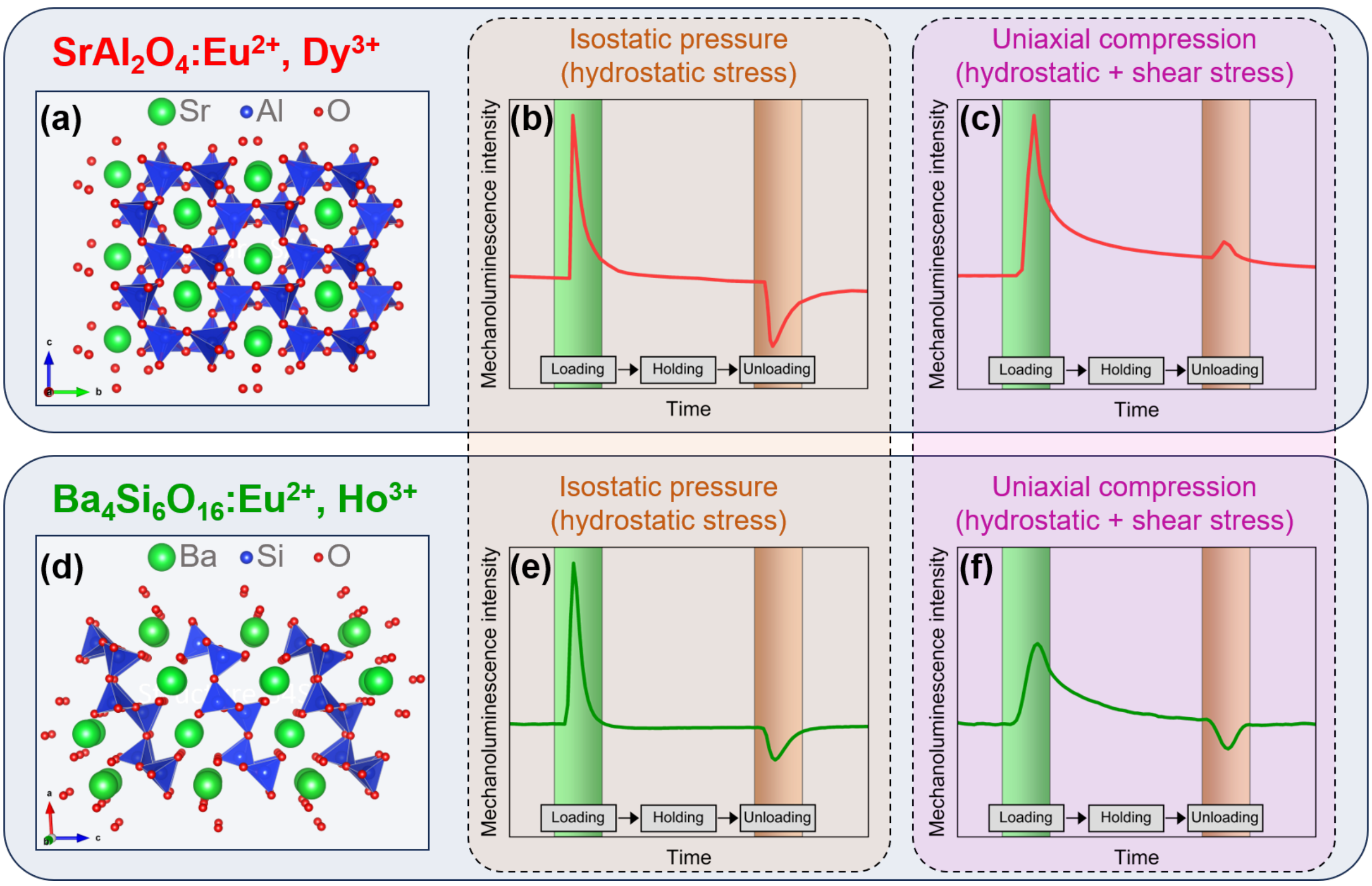

**Figure 1.** (a) Crystal structure of monoclinic and non-centrosymmetric $SrAl_2O_4$.[31] (b) Mechanoluminescence response of $SrAl_2O_4$:$Eu^{2+}$, $Dy^{3+}$ under hydrostatic pressure.[24,30] (c) Mechanoluminescence response of $SrAl_2O_4$:$Eu^{2+}$, $Dy^{3+}$ under combined hydrostatic and shear stresses. (a) Crystal structure of centrosymmetric $Ba_4Si_6O_{16}$.[32] (b) Mechanoluminescence response of $Ba_4Si_6O_{16}$:$Eu^{2+}$, $Ho^{3+}$ under hydrostatic pressure.[8,28,29] (c) Mechanoluminescence response of $Ba_4Si_6O_{16}$:$Eu^{2+}$, $Ho^{3+}$ under combined hydrostatic and shear stresses.

## 2. Computational methods

### 2.1 *DFT* calculations

The *DFT* calculations on the $Ba_4Si_6O_{16}$ (a = 12.477 Å, b = 4.685 Å, c = 13.944 Å and β = 93.54°, space group $P2_1$)[32] and $SrAl_2O_4$ (a = 8.447 Å, b = 8.816 Å, c = 5.163 Å and β = 93.42°, space group $P2_1$)[33] crystals were performed using the Vienna Ab initio Simulation Package (*VASP*) [34,35]. This code allows to determine the ground state properties of a given compound via *DFT*, employing on a plane-wave based Projector Augmented Wave (*PAW*) formalism. Three exchange-correlation functionals were considered: PBE,[36] a widely used generalized-gradient approximation (GGA); PBEsol,[37] a GGA reparametrized for densely packed solids; and HSE,[38] a screened hybrid functional that incorporates a fraction of exact (Hartree-Fock) exchange. The 4f levels of the rare-earth dopants (Eu, Ho, and Dy) must be treated explicitly, *i.e.*, not frozen in the pseudopotential. Because standard semi-local functionals such as *PBE* fail to adequately describe the strong on-site Coulomb repulsion of these localized 4f states, a Hubbard correction was applied within the rotationally invariant *DFT+U* formalism of Dudarev et al. An effective Hubbard parameter of $U_{eff} = U - J = 6$ eV was used for the 4f states of Eu, Ho, and Dy on top of the *PBE* functional. This same $U_{eff}$ value was applied consistently for the calculation of charge-transition levels of the three dopants, as well as for the evaluation of absorption and emission energies associated with the 4f-5d transitions of Eu. Calculations were carried out on a (a, 3b, c) supercell, both for defect-free and defect-containing configurations. The wavefunctions were expanded in plane-wave basis up to a kinetic energy cutoff of 500 eV, and the k-point sampling was chosen sufficiently fine to ensure the numerical convergence of all the calculated properties. In particular, the accurate calculation of the total energy was carried out using a 6×6×6 k-point mesh. The geometry was optimized until the forces on all the atoms were smaller than $3\cdot10^{-2}$ eV Å$^{-1}$. Total density of states (*TDOS*) and crystal orbital bond indices (*COBI*) were estimated based on *VASP* calculations using *PBE* functional and *LOBSTER* package.[39]

## 2.2 Charge-transition levels

The two methodological cornerstones of this work, the determination of thermodynamic charge-transition levels of point defects, and the estimation of neutral 4f → 5d excitation energies of $Eu^{2+}$, are summarized schematically in **Figure 2**.

Several intrinsic point defects in the $Ba_4Si_6O_{16}$ crystal were investigated, including cation and anion vacancies, antisites, and more complex defect configurations. The formation energy $E^f(D^q)$ of a defect $D$ with an effective charge state $q$ is defined as [20]:

$$E^f(D^q) = E_{tot}(D^q) - E_{tot}(bulk) - \sum_i n_i \mu_i + q(E_F + \varepsilon_{VBM}) + \Delta^q \quad (1)$$

where $E_{tot}(D^q)$ and $E_{tot}(bulk)$ are the total energies of the defect-containing and the defect-free supercells respectively, $n_i$ is the number of atoms with chemical potential $\mu_i$ added ($n_i >$ 0) or removed ($n_i < 0$) to form the defect, $E_F$ is the Fermi level, $\varepsilon_{VBM}$ is the valence band maximum, and $\Delta^q$ is a correction term accounting for spurious electrostatic interactions between periodically repeated defect images.

The charge-transition level $\varepsilon_{(q/q')}$, at which the effective charge state of a defect changes from $q$ to $q'$, corresponds to the Fermi-level position at which the formation energies $E^f(D^q)$ and $E^f(D^{q'})$ are equal:

$$\varepsilon_{(q/q')} = \frac{E^f(D^q; E_F = 0) - E^f(D^{q'}; E_F = 0)}{q' - q} \quad (2)$$

Here, $E^f(D^q; E_F = 0)$ and $E^f(D^{q'}; E_F = 0)$ are the formation energies of a defect $D$ in effective charge states $q$ or $q'$, respectively, evaluated with the Fermi level at the valence band maximum. The construction is illustrated schematically in **Figures 2a and 2b**. A point defect introduces one or more charge-transition levels (traps) within the band gap; the levels $\varepsilon_{(2+/1+)}$ and $\varepsilon_{(1+/0)}$ are shown relative to the valence-band (*VB*) and the conduction-band (*CB*), thus defining their trap depths. For each charge state q, the formation energy varies linearly with the Fermi level $E_F$, with a slope equal to q (Eq. 1); the transition levels ε(q/q′) appear as the Fermi-level positions at which two such lines cross. Plotting only the lowest-energy charge state at each value of $E_F$ yields the thick envelope of **Figure 2(b)**, whose successive segments correspond to the thermodynamically stable charge states of the defect across the band gap, and whose kinks define the charge-transition levels. Each transition level is associated with the trapping or detrapping of one or more charge carriers (*i.e.*, electrons $e^-$ and/or holes $h^+$), and can therefore be aligned within the host band gap to identify candidate trap states involved in luminescence processes. The point defects studied in this work were generated using the *PYDEFECT* code [40].

The Ba-O-Si chemical potential diagram, which delineates the stability domain of each phase in the Ba-O-Si system, was built from the formation enthalpy $\Delta H_f$ and the total energy $E_{tot}$ of each $\mathrm{Ba}_a\mathrm{Si}_b\mathrm{O}_c$ phase ($a + b + c = 1$). These quantities were obtained from relaxed atomic structures, with initial configurations sourced from the Materials Project database [41]:

$$\Delta H_f(\mathrm{Ba}_a\mathrm{Si}_b\mathrm{O}_c) = E_{tot}(\mathrm{Ba}_a\mathrm{Si}_b\mathrm{O}_c) - \sum_i \mu_i n_i \quad (3)$$

The chemical potentials $\mu_i$ entering equations (1) and (3) are the energies of the atomic reservoirs with which the crystal exchanges species during synthesis. Each is referenced to the *DFT* total energy corresponding elemental in its standard state, $\mu_i^0$, through the deviation $\Delta\mu_i = \mu_i - \mu_i^0$, which is the quantity that encodes the experimental growth conditions: an oxidizing or reducing atmosphere corresponds to a high or low $\Delta\mu_O$, respectively, while the chosen cation sources fix the remaining $\Delta\mu_i$ . Because the defect formation energy of Eq. (1) depends explicitly on the $\mu_i$, the synthesis conditions directly govern which point defects form and at what concentrations. The deviations are not free: each is bounded above by zero (**$\Delta\mu_i \leq 0$**, since $\Delta\mu_i > 0$ would precipitate the pure element), and equilibrium of the host phase requires that the stoichiometrically weighted deviations equal its Gibbs free energy of formation, $\Sigma_i n_i \Delta\mu_i = \Delta G_f(\text{host})$. At the 0 K DFT level, $\Delta G_f$ is approximated by the formation enthalpy $\Delta H_f$ of Eq. (3); the temperature and partial-pressure dependence of the synthesis atmosphere then enter chiefly through $\mu_O(T, p_{O_2})$. Competing binary and ternary phases impose further bounds on the individual $\Delta\mu_i$, and together these constraints delimit the host stability domain.

For oxygen, however, the standard *PBE* functional systematically underestimates the formation enthalpies of transition-metal and post-transition-metal oxides due to the overbinding of the $O_2$ molecule and the self-interaction error in the O 2*p* states. Following the correction scheme proposed by Wang *et al.* [42], a rigid energy shift of +1.36 eV is applied to the total energy of the $O_2$ molecule:

$$E_{O_2}^{corr} = E_{O_2}^{PBE} + 1.36 \quad (4)$$

This correction brings the calculated oxide formation enthalpies into agreement with experimental values from the JANAF thermochemical tables, and has become standard practice in defect calculations involving oxide hosts [40,43]. In practice, the correction shifts the oxygen chemical potential reference by +0.68 eV per O atom, which propagates into all formation enthalpies $\Delta H_f$ and into the chemical potential limits of the phase diagram. For the rare-earth impurity phases used to bound the RE chemical potentials ($RE_2O_3$ for Ho and Dy, and EuO or $Eu_2O_3$ for Eu depending on the synthesis atmosphere), the corresponding elemental reference energies were derived self-consistently from the GGA and GGA+U total energies of the respective oxides under the corrected oxygen reference, ensuring that each impurity phase sits exactly on the convex hull and that the chemical potential diagram remains internally consistent. It should be noted that, while the formation energy depends on the synthesis conditions, *i.e.*, chemical potentials of the individual constituents, the charge-transition levels (trap depths) do not.

### 2.3 Calculation of absorption/emission energy from *DFT* calculations

The *4f → 5d* neutral excitations of the $Eu^{2+}$ ion were estimated using constrained density functional theory (cDFT) and ΔSCF analysis of the total energies, following the approach of Jia *et al.* for a series of $Eu^{2+}$ phosphors.[44] The procedure is illustrated schematically in **Figure 2(c)**.

The principle of the method is to construct two distinct self-consistent solutions of the Kohn-Sham equations for the same $Eu^{2+}$-doped supercell, one with the activator in its $4f^7$ ground-state (*GS*) configuration, and one with the activator in its $4f^6 5d^1$ excited-state (*ES*) configuration, and to extract absorption and emission energies as differences of total energies between these two converged solutions, evaluated at appropriately chosen atomic geometries. Concretely, four single-point energies are computed: (i) $E_{Q_0}$, the total energy of the system in the *GS* electronic configuration at the *GS*-relaxed geometry $Q_0$; (ii) $E^*_{Q_0}$, the total energy of the system in the *ES* electronic configuration at the same *GS*-relaxed geometry $Q_0$ (*i.e.*, the geometry has not yet relaxed in response to the electronic excitation); (iii) $E_{Q_1}$, the total energy of the system in the *ES* electronic configuration at the *ES*-relaxed geometry $Q_1$, obtained by relaxing the supercell with the $4f^6 5d^1$ occupation imposed; (iv) $E^*_{Q_1}$, the total energy of the system in the *GS* electronic configuration at the *ES*-relaxed geometry $Q_1$.

The absorption energy then corresponds to the vertical (Franck-Condon) transition at fixed nuclear positions $Q_0$, and the emission energy to the analogous vertical transition at $Q_1$:

$$E_{abs} = E^*_{Q_0} - E_{Q_0} \tag{5}$$

$$E_{em} = E^*_{Q_1} - E_{Q_1} \tag{6}$$

This four-point construction is shown graphically in **Figure 2(c)** on the *GS* (purple) and *ES* (green) Born-Oppenheimer surfaces, parameterized by a representative configuration coordinate. The $4f^6 5d^1$ electronic configuration was imposed by constraining the occupation of the relevant Eu-centered orbitals during the SCF cycle, as implemented in cDFT. Convergence difficulties of the type recently reported recently by Y. Xiong and G. Hautier[45] were encountered when applying this procedure within recent VASP releases. These were resolved by reverting to an earlier version of VASP in conjunction with the patch proposed by Lindvall et al.,[46] which addresses this known issue.

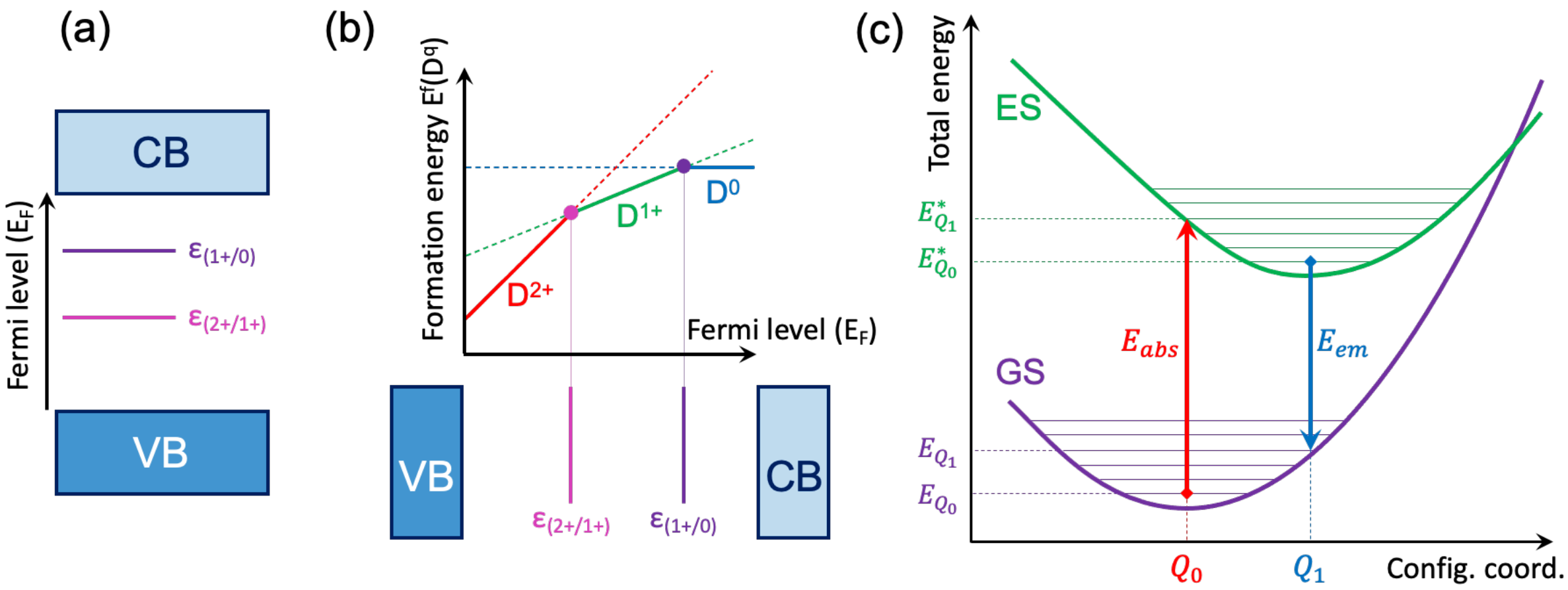

**Figure 2.** Schematic representation of the two methodologies used in this work to extract defect-related quantities from first-principles total-energy calculations. (a) A point defect introduces one or more charge-transition levels (traps) within the band gap; the levels $\varepsilon_{(2+/1+)}$ and $\varepsilon_{(1+/0)}$ are shown relative to the valence-band (*VB*) and the conduction-band (*CB*), thus defining their trap depths. (b) These charge-transition levels are obtained from the Fermi-level dependence of the defect formation energy $E^f$. The lines show $E^f(D^q)$ for charge states q = 2+, 1+, and 0 (red, green, and blue, respectively), each with a slope equal to q (Eq. 1). The lower envelope traces the more stable charge state at every Fermi level; its kinks define the charge-transition levels $\varepsilon_{(2+/1+)}$ and $\varepsilon_{(1+/0)}$ (Eq. 2), aligned below with respect to the *VB* and *CB*. (c) Estimation of the absorption ($E_{abs}$) and emission ($E_{em}$) energies of a localized neutral excitation, here, the 4f → 5d transition of $Eu^{2+}$, using constrained DFT (cDFT) combined with ΔSCF total-energy differences. The ground-state (*GS*, purple) and excited-state (*ES*, green) Born-Oppenheimer surfaces are shown as a function of a representative configuration coordinate Q. $E_{abs}$ corresponds to a vertical (Franck-Condon) transition from the ground-state minimum $Q_0$ to the excited-state surface; $E_{em}$ corresponds to the reverse vertical transition from the relaxed excited-state minimum $Q_1$.

### 2.4 Application of hydrostatic and shear stresses

Hydrostatic stress was applied to the defect-containing and defect-free supercells by controlling the isostatic pressure. Negative and positive isostatic pressures allowed to study compressive and tensile stresses, respectively. Shear stress was generated by inducing the displacement of a plane along a given axis. This caused an angular deformation of the angle between (i) the axis along which the shear stress $\tau$ was generated, and (ii) the axis perpendicular to the considered plane (see Erreur ! Source du renvoi introuvable.**(a-b)**).

## 3. Results and discussion

### 3.1 Delimitation of the system's conditions

To shed light on the energy levels active in the PersL of $Ba_4Si_6O_{16}$:$Eu^{2+}$, $Ho^{3+}$, several point defects within this crystal were considered. The atomic structure of $Ba_4Si_6O_{16}$ (**Figure 3(a)**) features two inequivalent barium sites ($Ba_1$, $Ba_2$) and three inequivalent silicon sites ($Si_1$, $Si_2$, $Si_3$), while the eight inequivalent oxygen positions split into two distinct families based on their local coordination: four *inner* oxygens ($O_{in}$: $O_1$–$O_4$), each bridging two $SiO_4$ tetrahedra, and four *outer* oxygens ($O_{out}$: $O_5$–$O_8$), each bonded to a single silicon neighbor. This crystallographic inequivalence gives rise to a rich set of candidate defect sites (*e.g.*, substitutional Eu and Ho on either Ba site, and oxygen vacancies on either $O_{in}$ or $O_{out}$ positions) whose formation energies and electronic levels depend sensitively on the synthesis conditions.

The thermodynamic accessibility of each defect is governed by the chemical potentials of the constituent elements, which are themselves constrained by the stability of $Ba_4Si_6O_{16}$ against decomposition into competing binary and ternary phases of the Ba–Si–O system. The full three-dimensional chemical-potential phase diagram is shown in **Figure 3(b)**, where the stability domain of $Ba_4Si_6O_{16}$ appears as a narrow polyhedron bounded by $SiO_2$, $BaSi_2O_5$, $BaSiO_3$, $Ba_2SiO_4$, and $Ba_3Si_4$. To analyze this domain more quantitatively, we project it onto the ($\Delta\mu_{Ba}$, $\Delta\mu_{Si}$) plane, as shown in **Figure 4**. This projected domain is particularly narrow: the points A and B, as well as C and D, lie close to one another (less than 0.02 eV apart), so we focus on the A – C line in what follows.

Point A corresponds to oxygen-rich conditions ($\mu_O$ = 0 eV), whereas point C corresponds to oxygen-poor conditions ($\mu_O$ = -4.95 eV). In addition, the green dashed line and the red dotted line in **Figure 3** indicate the oxygen chemical potentials corresponding, respectively, to $H_2/N_2$ ($\Delta\mu_O = -3.08$ eV) and air ($\Delta\mu_O = -1.96$ eV) synthesis atmospheres at 1573 K, as determined from thermodynamic data using the JANAF tables. These represent conditions intermediate between points A and C, while pure $N_2$ corresponds to $\Delta\mu_O = -4.95$ eV, essentially coinciding with point C. Using the *PBE* functional, we computed the charge-transition levels of the oxygen vacancies $V_O^q$ for the 8 inequivalent oxygen sites of $Ba_4Si_6O_{16}$ under chemical potential conditions A and C. These are shown in **Figure 5(a)** as faded and fully opaque lines, respectively. The experimental conditions obviously have a strong incidence on the formation energies of oxygen vacancies, which decrease from about 6 to 1 eV as $\mu_O$ decreases from 0 to -4.95 eV (*i.e.*, as conditions become increasingly oxygen-poor). Since previous studies reported the synthesis of the $Ba_4Si_6O_{16}$:$Eu^{2+}$, $Ho^{3+}$-containing glass-ceramic and of the $Ba_4Si_6O_{16}$:$Eu^{2+}$, $Ho^{3+}$ crystal under reducing conditions ($N_2$ atmosphere in the presence of $Si_3N_4$ [8] and $N_2$:$H_2$ atmosphere [28], respectively), point C represents the practical systems more accurately. The calculations discussed in the following were therefore carried out under oxygen-poor conditions.

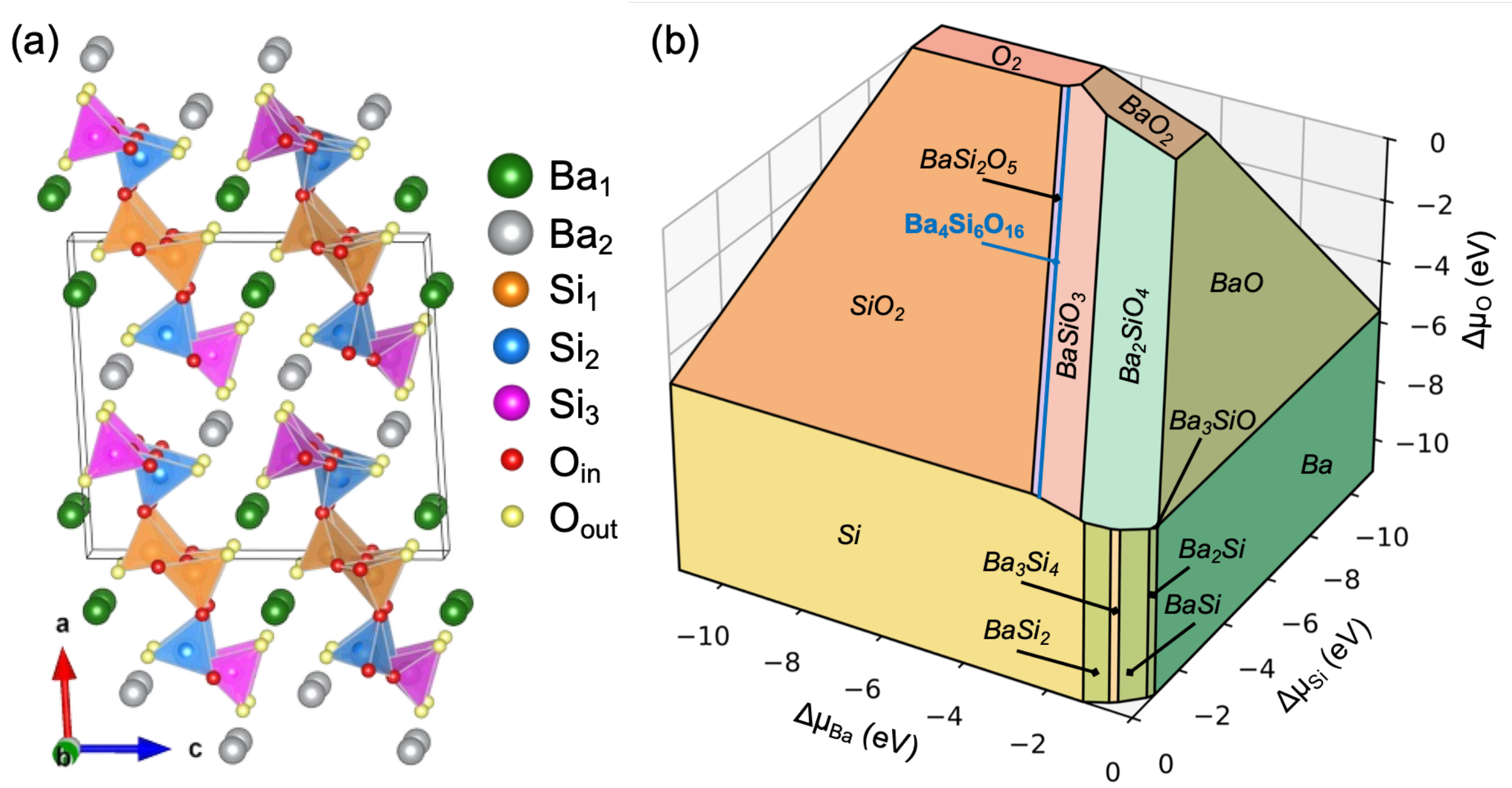

**Figure 3.** (a) Schematic representation of the atomic structure of $Ba_4Si_6O_{16}$. Two inequivalent barium sites ($Ba_1$, $Ba_2$) and three inequivalent silicon sites ($Si_1$, $Si_2$, $Si_3$) are highlighted. The eight inequivalent oxygen positions are categorized into two groups according to their local coordination: inner oxygen atoms $O_{in}$ ($O_1$–$O_4$), shown in red, which bridge two silicon neighbors, and outer oxygen atoms $O_{out}$ ($O_5$–$O_8$), shown in yellow, which are bonded to a single silicon neighbor. $SiO_4$ tetrahedra are colored according to the central Si site ($Si_1$ orange, $Si_2$ blue, $Si_3$ magenta), and the unit cell is outlined in black. (b) Chemical potential phase diagram of the Ba–Si–O ternary system, obtained from our *PBE* calculations with *PYDEFECT*, showing the stability region of $Ba_4Si_6O_{16}$ (blue) bounded by competing phases as a function of the chemical potentials $\Delta\mu_{Ba}$, $\Delta\mu_{Si}$, and $\Delta\mu_O$ (in eV) referenced to their standard elemental states.

Furthermore, calculations of the charge-transition levels of oxygen vacancies were also performed with two additional functionals to ensure the reliability of the chosen method, namely *PBEsol* and *HSE*. The *PBE* (**Figure S3(a)**) and *PBEsol* (**Figure S3(b)**) calculations yield close values of the charge-transition level positions and their associated formation energies. The charge-transition levels can be divided into two categories: those arising from vacancies at bridging oxygen sites (*i.e.*, oxygen atoms bonded to two silicon atoms), which lie closer to the valence band, and those arising from vacancies at non-bridging oxygen sites (*i.e.*, oxygen atoms bonded to only one silicon atom), which lie closer to the conduction band. The calculated band gap is 4.51 eV (Erreur ! Source du renvoi introuvable.), close to the 4.6 eV value calculated by Li *et al.* (4.6 eV) [47], and to the experimental value of 5.15 eV reported by Zhao *et al.* [48]. In contrast, the results obtained with the *HSE* functional (**Figure S3(c)**) differ substantially from those of the two other functionals, with a significantly overestimated band gap (6.22 eV) and charge-transition levels located deeper in the band gap. Moreover, calculations of the charged defects were found difficult to converge in *HSE*. These numerical instabilities led us to give greater credence to the results obtained with the *PBE* and *PBEsol* functionals. Accordingly, the results discussed below were obtained using the *PBE* functional. While the *PBE* functional typically underestimates the band gap, the discrepancy here is small enough that it does not significantly affect the results. The projected densities of states (*PDOS*) allow us to identify the main character of the valence band (*VB*) and conduction band (*CB*), which correspond to O(2p) and Ba(5d) states, respectively. The top of the *VB* (from 0 to -2 eV) mainly consists of non-bonding O(2p) states, while its lower part (from -2 to -5 eV) includes O(2p) states that interact covalently with Si(3p) states.

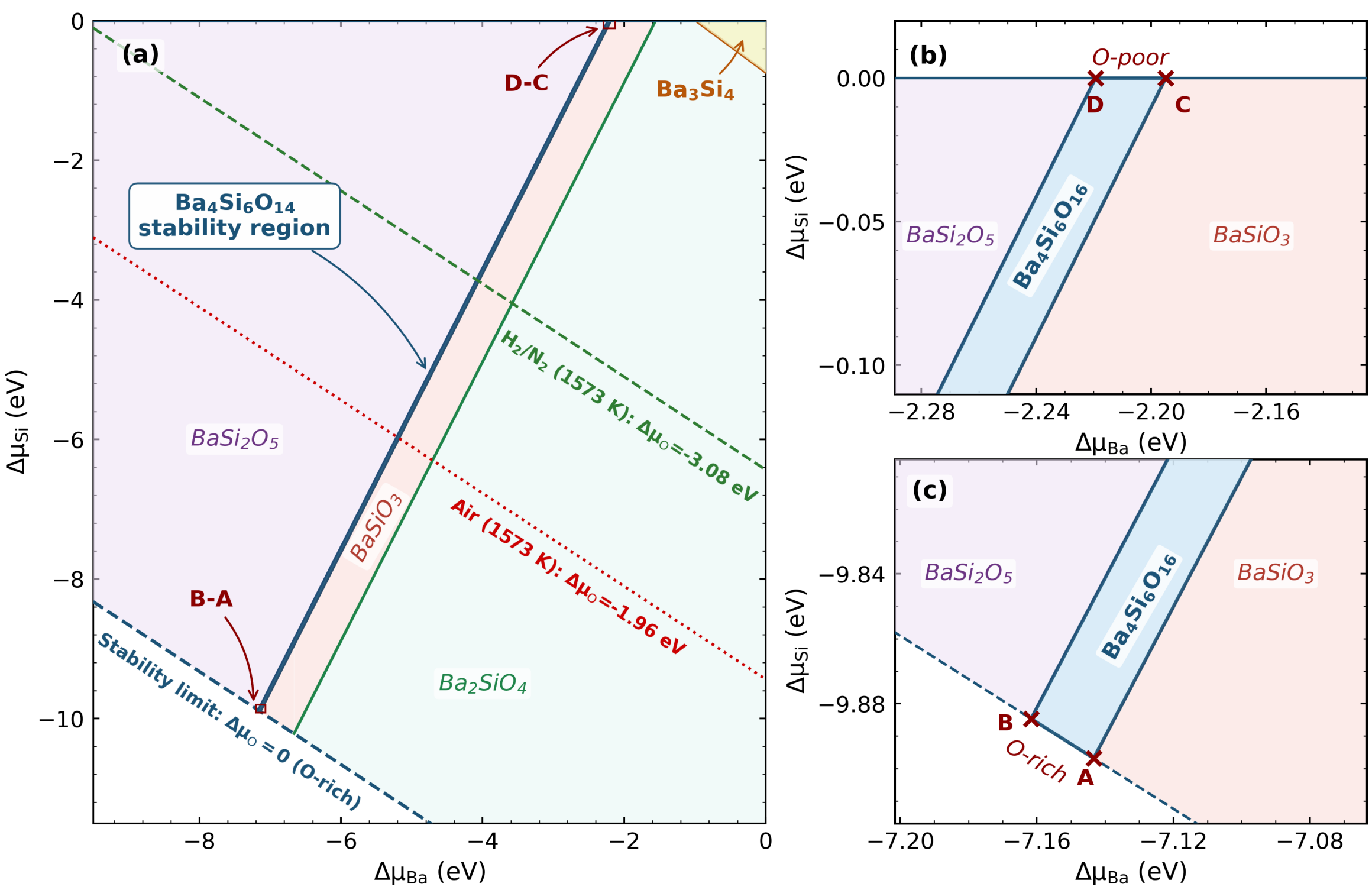

**Figure 4.** Two-dimensional projection of the Ba–Si–O chemical-potential phase diagram onto the ($\Delta\mu_{Ba}$, $\Delta\mu_{Si}$) plane. The domain of stability of the $Ba_4Si_6O_{16}$ phase, shown in blue, is bounded by the vertices A, B, C, and D, along which the oxygen chemical potential $\Delta\mu_O$ evolves continuously from the oxygen-rich limit (vertices A-B) to the oxygen-poor limit (vertices C – D). The blue dashed line represents the stability limit at $\Delta\mu_O = 0$ (O-rich), defined by the formation enthalpy of $Ba_4Si_6O_{16}$. The competing phases $BaSi_2O_5$, $BaSiO_3$, $Ba_2SiO_4$, and $Ba_3Si_4$ are also shown. The green dashed line and the red dotted line indicate the oxygen chemical potential corresponding, respectively, to $H_2/N_2$ ($\Delta\mu_O$ = -3.08 eV) and air ($\Delta\mu O$ = -1.96 eV) synthesis atmospheres at 1573 K, as determined from thermodynamic data using the JANAF tables. Both experimental conditions fall within the stability domain of $Ba_4Si_6O_{16}$ with $H_2/N_2$ approaching the oxygen-poor limit (vertices C – D) and air lying closer to the oxygen-rich boundary (vertices A – B).

### 3.2 Active point defects and charge carriers in the PersL mechanism

To identify the active point defects and charge carriers within the PersL mechanism of $Ba_4Si_6O_{16}:Eu^{2+}$, $Ho^{3+}$, both intrinsic defects (cationic and anionic vacancies, anion-for-cation antisite substitutions) and complex defects (adjacent vacancy pairs) are considered. Their corresponding charge-transition levels $\varepsilon(q/q')$ and formation energies $E^f$ are shown in **Figure**

**5(a-h)**. Based on these results, several defect types can be confidently excluded from further consideration for the following reasons.

(i) Barium vacancies $V_{Ba}^{q}$ (**Table S1, panel b**) exhibit formation energies of 2.1 – 2.9 eV under O-rich (oxidizing) conditions and 7.1 – 7.8 eV under O-poor (reducing) conditions, with shallow first acceptor transitions ε(0/−1) at only 0.16 – 0.20 eV above the valence band maximum (*VBM*). The second acceptor ε(−1/−2) is deeper, at 0.58 – 0.72 eV. Despite the shallow first level, the formation energies remain too high for any significant equilibrium concentration of $V_{Ba}$ under either chemistry. Silicon vacancies $V_{Si}^{q}$ (**Table S1, panel c**) are even less favorable, with $E^f$ spanning 9.2 – 14.5 eV at the O-poor limit and 4.2 – 4.6 eV at the O-rich limit, and four sequential acceptor transitions ε(0/−1)…ε(−3/−4) distributed between $E_F$ = 0.46 and 3.01 eV; site 2 shows a negative-U ε(−2/−4) skipping the −3 state. Both defect types are therefore irrelevant to the PersL mechanism. The $V_O$–$V_{Ba}$ complex (**Table S1, panel h**) has a formation energy of 6.63 eV that is strictly independent of position along the oxygen partial pressure, a direct consequence of the exact cancellation of $\Delta\mu_O$ and $\Delta\mu_{Ba}$ across the $Ba_4Si_6O_{16}$ stability field, with a negative-U transition ε(0/−2) at $E_F$ = 1.30 eV, which is energetically inaccessible.

(ii) Silicon antisite defects $Si_O^q$ (**Table S1, panel d**) are eliminated from the PersL mechanism on electronic rather than thermodynamic grounds. Under O-poor conditions the highly charged $Si_{O7}^{+6}$ and $Si_{O7}^{+4}$ states even reach negative formation energies ($E^f$ = −3.33 and −2.94 eV at low $E_F$), so $Si_O$ is in principle accessible. However, every transition level ε(q/q′) lies between 0.25 and 3.38 eV above the *VBM*, and most are negative-U, the +4 state collapses to +2, and at sites 7 and 8 the +6 collapses directly to +4 without a stable intermediate. None of these levels is shallow enough to trap and re-emit a carrier at energies relevant to the $Eu^{2+}$ emission channel, so $Si_O$ cannot participate in the long-lived PersL process.

(iii) Rare-earth substitution $RE_{Ba}^{q}$ (**Table S1, panels e-g**) is explored for Eu, Ho, and Dy (**Figure 5(e-g)**). Eu substitution under O-poor (reducing) conditions, referenced to EuO, leads to spontaneous incorporation with $E^f$ = −0.68 and −0.79 eV at the two Ba sites; under O-rich conditions, referenced to $Eu_2O_3$, formation is still favorable but no longer exothermic ($E^f$ = 1.32 and 1.21 eV). The (+1/0) transition level lies at $E_F$ = 1.39 – 1.42 eV above the *VBM*, *i.e.*, in the lower half of the gap, so $Eu^{2+}$ is the stable charge state across the full range of Fermi-level positions accessible under typical synthesis conditions, fully consistent with the experimentally observed $Eu^{2+}$ luminescence (broadband emission, in contrast to $RE^{3+}$ sharp emissions). $Ho^{3+}$ and $Dy^{3+}$ substitutions show moderate formation energies under O-poor conditions (1.11 – 1.23 eV for Ho, 0.87 – 1.01 eV for Dy), confirming that both are readily incorporated into the host lattice. Their (+1/0) transition levels sit at $E_F$ = 1.22 eV (Ho) and 1.04 (Dy) 1.04 eV below the conduction band minimum (*CBM*). These depths are markedly larder than the ~0.5-0.7 eV energy window typically associated with room-temperature persistent luminescence in $Eu^{2+}$-activated phosphors [13]. For instance, the dominant Dy-related electron trap in $SrAl_2O_4$:$Eu^{2+}$, $Dy^{3+}$ that produces hours-long afterglow lies at approximately 0.65 eV below the *CBM* [49]. Within the simple thermal-detrapping picture (see below), the $Ho^{3+}$ and $Dy^{3+}$ traps in $Ba_4Si_6O_{16}$ would release carriers on timescales of months to years at room temperature, far too slowly to contribute to the experimentally observed afterglow. $Ho^{3+}$ and $Dy^{3+}$ substitutional defects are therefore unlikely to be the primary carrier reservoirs governing the PersL kinetics in this system. Consistently, $Eu^{2+}$ alone is sufficient to observe both afterglow and mechanoluminescence in $Ba_4Si_6O_{16}$:$Eu^{2+}$, although controlled $RE^{3+}$ co-doping enhances both intensities. We therefore interpret the $RE^{3+}$ co-dopants as auxiliary deep reservoirs that increase the long-term storage capacity of the host, possibly contributing to ultra-long-lived components of the afterglow not resolved on standard experimental timescales, without driving the primary trapping-detrapping mechanism in $Ba_4Si_6O_{16}$:$Eu^{2+}$, $RE^{3+}$.

As a consequence, the only relevant defects exhibiting accessible formation energies and transition levels at adequate positions relative to a band edge are the oxygen vacancies $V_O^q$ (**Table S1, panel a**), which are conventionally considered electron traps in PersL mechanisms. Under O-poor conditions, all eight crystallographically inequivalent $V_O$ sites have formation energies $E^f < 1.2$ eV at their respective transition levels (range 0.27 – 1.18 eV). Six sites ($V_{O1}$, $V_{O2}$, $V_{O3}$, $V_{O4}$, $V_{O7}$, $V_{O8}$) are negative-U defects exhibiting direct (+2/0) transitions without a stable +1 intermediate, with levels spanning 0.43 – 2.16 eV above the *VBM*. The two remaining sites $V_{O5}$ and $V_{O6}$ show sequential (+2/+1/0) transitions at 1.08/1.91 eV and 1.57/1.97 eV respectively. In order to assess which of these levels can act as PersL-relevant traps at room temperature, we estimate the thermal-detrapping lifetime using a first-order Arrhenius expression, $\tau = 1/s \exp(E_t/k_BT)$, with attempt frequency $s \approx 10^{12}$ $s^{-1}$ and T = 300 K. At this attempt frequency, useful afterglow on timescales from seconds to hours corresponds to trap depths $E_t \approx 0.7$–0.95 eV from the relevant band edge; depths of 1.0–1.15 eV give lifetimes from hours to roughly one month; while levels $\gtrsim$ 1.3 eV release carriers on timescales of centuries or longer, far too slowly to contribute to PersL at room temperature.

Considered as electron traps, all $V_O$ levels lie 2.3 – 4.1 eV below the *CBM* and are therefore effectively frozen at room temperature, they cannot release captured electrons on any timescale relevant to afterglow. The conventional $V_O$ electron-trap picture is incompatible with these results. The $Ho^{3+}$ and $Dy^{3+}$ ε(+1/0) levels (1.22 and 1.04 eV below the *CBM,* respectively) lie shallower than the $V_O$ electron levels but, as discussed above, are still too deep to act as primary electron traps governing the experimentally observed room-T afterglow on time scale of seconds to hours.

The picture changes when $V_O$ is considered as a *hole trap*. The four bridging-oxygen sites $V_{O1}$–$V_{O4}$ have ε(+2/0) levels at 0.43, 0.57, 0.70, and 0.87 eV above the *VBM*, placing $V_{O3}$ and $V_{O4}$ squarely in the room-T PersL window (lifetimes of ~1 s and ~3 min respectively); $V_{O5}$ and $V_{O6}$ contribute additional levels at 1.08 and 1.57 eV (hours to decades). $V_{O1}$ and $V_{O2}$ are too shallow to retain a hole at room temperature. This is a defining feature of the $Ba_4Si_6O_{16}$ system: in most oxide PersL phosphors $V_O$ acts as an electron trap because the +2 → +1 → 0 reduction populates a defect level just below the *CBM*, but here the negative-U ε(+2/0) collapse pushes the level deep into the lower half of the gap, near the *VBM*. The same near-*VBM* positioning of $V_O^q$ is documented in amorphous $SiO_2$ [50,51], where the negative-U behavior originates from a structural rearrangement of the neutral vacancy into a Si–Si bond, and hole capture on $V_O^0$ produces a small-formation-energy $V_O^{1+}$ state [50] (let us recall that holes are mobile, albeit less than electrons [19,52]). The corner-sharing $SiO_4$ tetrahedral chains in $Ba_4Si_6O_{16}$, separated by $BaO_8$ polyhedral sheets, provide exactly the local flexibility required for the same Si–Si rebonding to occur at the bridging $O_{in}$ sites. The PersL mechanism in this material is therefore driven by hole trapping at the $V_{O3}$ and $V_{O4}$ bridging-oxygen vacancies, with electron storage on the $RE^{3+}$ co-dopants, a mechanism enabled by the negative-U Si–Si rebonding inherited from the silicate framework, rather than by the simple $V_O^{1+}$ / $V_O^{2+}$ electron-trap chemistry of close-packed oxide hosts.

To quantify the formation of the Si-Si bond in $Ba_4Si_6O_{16}$, we have estimated the integrated crystal orbital bond index (*ICOBI*) [39]. This quantity provides a bond index for solid-state materials, analogous to those used for molecules: *ICOBI* values of 1, 2 and 3 correspond to single, double and triple covalent bonds, respectively, while *ICOBI* = 0 indicates that the interaction between the two probed atoms is not covalent, *i.e.*, there is no orbital overlap. In the defect-free host compound, there are no Si-Si bonds. Introducing a $V_O^0$ defect along the Si(1)-O-Si(2) path in a corner-sharing $SiO_4$ chain triggers a structural reorganization: the Si(1)-Si(2) contracts from ~3 Å to ~2.4 Å, and the *ICOBI* rises from ~0 and 0.70, evidencing the formation of a genuine Si–Si single bond (see Erreur ! Source du renvoi introuvable. and **S5**). This result is consistent with our formation-energy analysis. The four oxygen vacancy sites with transition levels within ~0.9 eV of the *VBM* ($V_{O1}$, $V_{O2}$, $V_{O3}$, $V_{O4}$) correspond exclusively to bridging oxygens ($O_{in}$) located at Si–O–Si corners in the silicate chain, where the two neighboring Si atoms can approach each other upon vacancy formation and form a Si–Si bond. By contrast, $V_{O5}$, $V_{O6}$, $V_{O7}$, $V_{O8}$ are terminal oxygens ($O_{out}$) bonded to a single Si atom, for which no such Si-Si rebonding is possible. Their transition levels accordingly lie deeper in the gap (1.08 to 2.16 eV above the *VBM*): $V_{O5}$ and $V_{O6}$ retain the standard sequential (+2/+1/0) progression with no negative-U behavior, and $V_{O7}$ and $V_{O8}$ do show negative-U (+2/0) collapse but with much higher kink positions than the bridging sites, reflecting weaker structural relaxation around the terminal-oxygen environment. Hole trapping at $V_O^0$, and the subsequent formation of $V_O^{2+}$, reverses this reorganization: the Si(1)-Si(2) distance increases back to ~4.5 Å and the *ICOBI* drops to ~0, thereby breaking the previously formed Si-Si bond. It is noteworthy that the Si(1)-Si(3) distance decreases to 2.55 Å, but no Si-Si bond is formed (*ICOBI* = 0.04). The structural reorganization around $V_O^0$ and $V_O^{2+}$ point defects, together with the corresponding Si(1)-Si(2) distances, is shown in **Figures 6(a) and 6(b)** respectively.

This picture is fully consistent with the negative-U character identified for six of the eight oxygen vacancy sites: the large structural relaxation accompanying the $V_O^0 \rightarrow V_O^{2+}$ transition (Si–Si bond formation and subsequent breaking) provides the lattice distortion energy that stabilizes the 2+ charge state relative to the +1 intermediate, making the +1 state thermodynamically inaccessible. The ε(2+/0) transition levels of bridging oxygen vacancies ($V_{O_{in}}$) lie within ~0.9 eV of the *VBM*, making them well-suited as room-temperature hole traps in the PersL cycle: at 300 K and with an attempt frequency of $\sim 10^{12}$ $s^{-1}$, the corresponding thermal-release lifetimes range from microseconds ($V_{O1}$) to several minutes ($V_{O4}$), spanning the timescales relevant to afterglow emission.

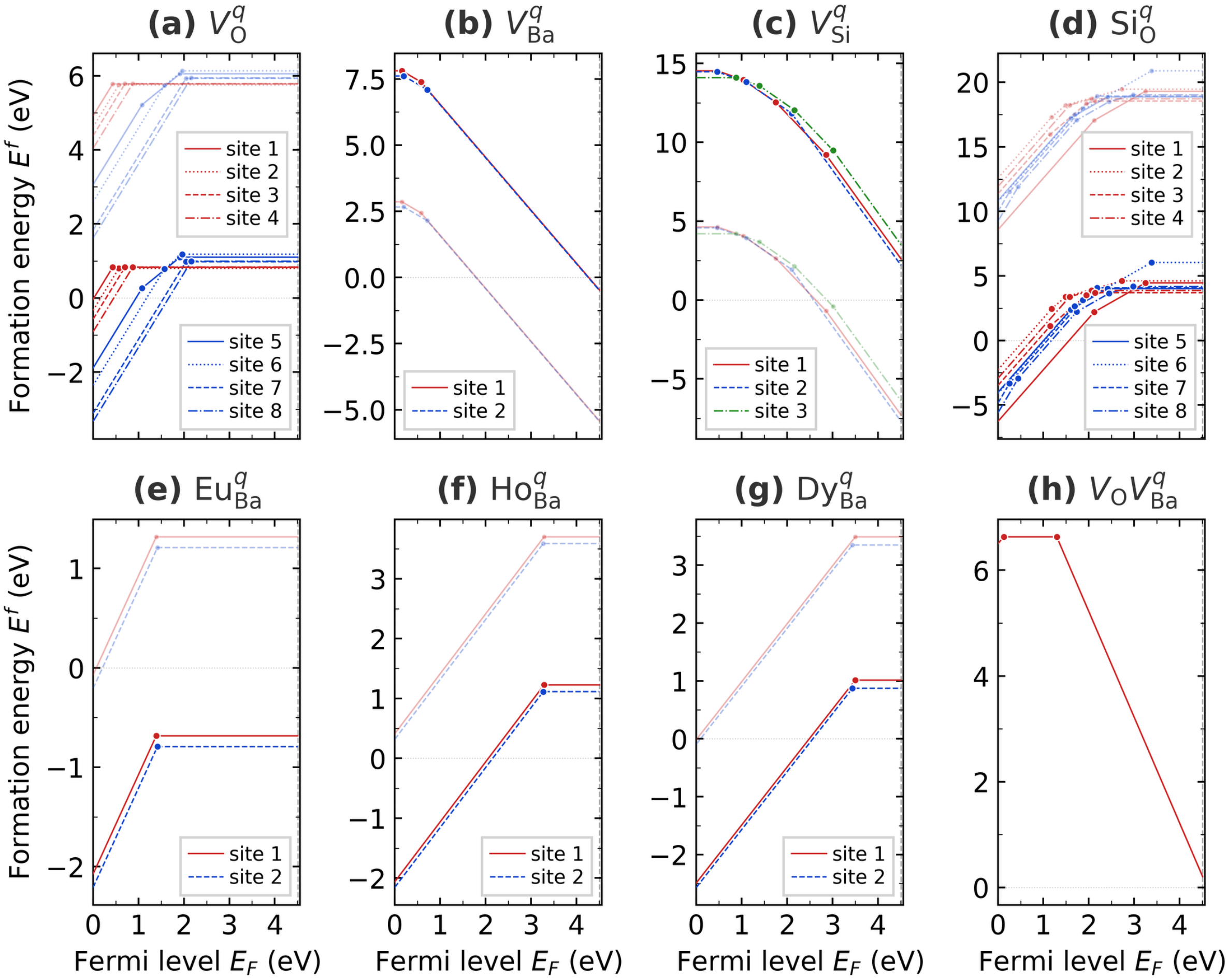

**Figure 5.** Formation energies $E^f$ of various point defects in $Ba_4Si_6O_{16}$ as a function of the Fermi level $E_F$, ranging from the valence band maximum (*VBM*) to the conduction band minimum (*CBM*), under two thermodynamic conditions at 1573 K: oxidizing atmosphere (condition A, O-rich, thin/transparent lines) and reducing atmosphere (condition C, O-poor, bold/opaque lines). For rare-earth substitutions, chemical potentials are referenced to $RE_2O_3$, except for the $Eu_{Ba}$panel under $H_2/N_2$ conditions, where EuO is used as the reference. For clarity, only the segments corresponding to the charge state $q$ with the lowest formation energy are shown. A positive slope indicates a positively charged defect, a negative slope a negatively charged defect, and a zero slope a neutral defect. Charge-transition levels $\varepsilon_{(q/q')}$ are marked by circular markers. Different line styles and colors distinguish inequivalent crystallographic sites within each defect family. (a) Oxygen vacancies $V_O^q$ (sites 1 – 8). (b) Barium vacancies $V_{Ba}^q$ (sites 1 – 2). (c) Silicon vacancies $V_{Si}^q$ (sites 1 – 3). (d) Silicon antisite on oxygen $Si_O^q$(sites 1 – 8). (e) Europium substituted on barium $Eu_{Ba}^q$ (sites 1 – 2). (f) Holmium substituted on barium $Ho_{Ba}^q$ (sites 1 – 2). (g) Dysprosium substituted on barium $Dy_{Ba}^q$ (sites 1 – 2). (h) Oxygen vacancy neighboring a barium vacancy $V_O$–$V_{Ba}^q$ (divacancy complex).

These calculations emphasize the crucial role of oxygen vacancies in the PersL mechanism of $Ba_4Si_6O_{16}$:$Eu^{2+}$, $Ho^{3+}$. Furthermore, the positioning of the corresponding charge-transition levels highlights the significant role of holes as charge carriers, a phenomenon previously observed in a-$SiO_2$. At room temperature, energy levels up to approximately 1 eV contributes to PersL intensity over a timescale of a few seconds to a few hours. Accordingly, only the charge-transition levels (i) $\varepsilon_{(2+/0)}$ located at 0.43, 0.57, 0.70 and 0.87 eV (associated with $V_{O_1}$, $V_{O_2}$, $V_{O_3}$ and $V_{O_4}$ respectively, all bridging $O_{in}$ sites along Si–O–Si corners), and (ii) $\varepsilon_{(2+/1+)}$ located at 1.08 eV (associated with $V_{O_5}$, a sequential (2+/1+/0) site) need to be considered. This distribution of trap depths is consistent with the experimentally observed Gaussian distribution in $Ba_4Si_6O_{16}:Eu^{2+}$, $Ho^{3+}$ (see ref. [28] and [29]), where approximately 90% of the observed levels fall between 0.43 and 1.08 eV. It is also consistent with the observation that mechanoluminescence behavior during loading and unloading is independent of the *RE* co-doping. In fact, *RE* co-doping, due to differences in ion size and valence between *RE* and $Ba^{2+}$, primarily affects the concentrations and associated charge-transition levels of neighboring point defects.[28]

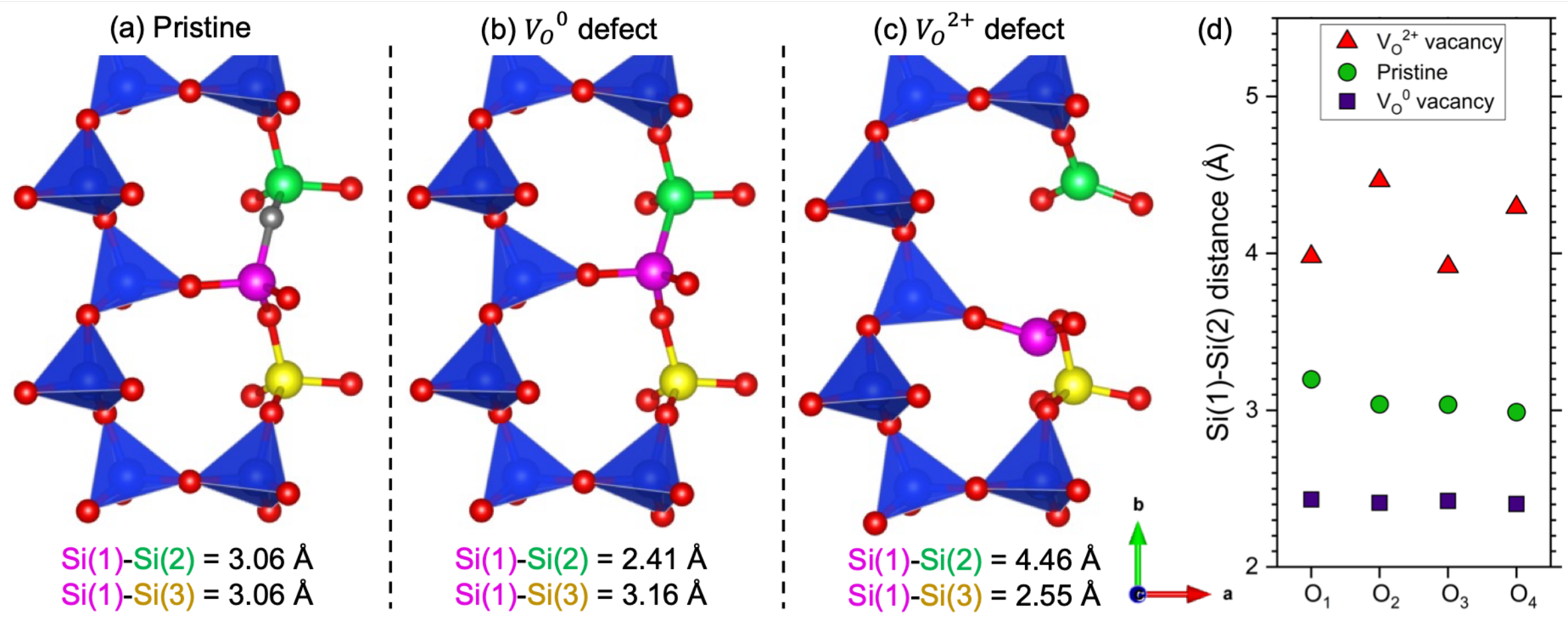

**Figure 6.** Structural reorganization in $Ba_4Si_6O_{16}$ induced by oxygen vacancy formation. (a) Pristine local structure showing the oxygen atom (grey) prior removal, with equal Si(1)–Si(2) and Si(1)–Si(3) distances of 3.06 Å. (b) Relaxed structure following the formation of neutral oxygen vacancy ($\boldsymbol{V_O^0}$), in which the Si(1)–Si(2) distance contracts to 2.41 Å, indicating the formation of a Si–Si bond across the vacancy site. (c) Relaxed structure following the formation of a doubly-charged oxygen vacancy ($\boldsymbol{V_O^{2+}}$), where the Si(1)–Si(2) distance expands to 4.46 Å as the silicon atoms relax outward. (d) Si(1)–Si(2) distances for vacancies created at four symmetry-distinct oxygen sites ($O_1$–$O_4$), comparing the pristine structure (green circles), $\boldsymbol{V_O^0}$ (purple squares), and $\boldsymbol{V_O^{2+}}$ (red triangles). Si(1), Si(2), and Si(3) atoms are shown in pink, green, and yellow, respectively; remaining $SiO_4$ tetrahedra are rendered in blue with oxygen atoms in red. Barium atoms are omitted for clarity.

### 3.3 Theoretical estimation of Eu-related optical transitions

To accurately define the processes involved in the PersL of $Ba_4Si_6O_{16}:Eu^{2+}$, we need to estimate the optical transitions from first principles and compare them with our experimental measurements. Similar to $SrAl_2O_4:Eu^{2+}$,[31] we observe two emission bands (blue and green), which are experimentally attributed to $Eu^{2+}$ ions occupying two inequivalent crystallographic sites.[28] To confirm this assignment, we have estimated the optical transitions (both absorption and emission processes of $Eu^{2+}$ 5d-4f) for $Ba_4Si_6O_{16}:Eu^{2+}$. We also examined the case of $SrAl_2O_4:Eu^{2+}$ to validate our theoretical approach. **Table 1** presents the results from our DFT+U calculations using cDFT, alongside experimental data and previous simulations for comparison. All theoretical data utilized cDFT calculations combined with ΔSCF analysis to extract the absorption (Abs) and emission (Em) transition energies. The best agreement between experiments and theory was achieved by Jia *et al.* with relative errors smaller than 10%.[44] In contrast, Hoang was unable to achieve a similar level of accuracy using DFT+U calculations with a computational setup akin to that of *Jia et al.*, resulting in relative deviations of up to 42%.[21] The only difference between these calculations lies in the computational code used: Jia *et al.* employed ABINIT while Hoang used VASP. However, both codes are based on plane-wave basis sets and pseudopotentials, so we do not anticipate such discrepancies. Indeed, when we conducted similar calculations using VASP, we encountered convergence issues, which we resolved with a patched version of an older version of the VASP code. Our results are now closer to those of Jia *et al.* with relative deviations smaller than 12%. It is also worth noting that Hoang reported *HSE* results that significantly overestimate the experimental optical transitions, with relative deviation exceeding 23%. This suggests that using *PBE* functional, along with a Hubbard correction for the 4f(Eu) states, yields more accurate results than the *HSE* approach.

**Table 1.** The calculated absorption and emission energies (in eV) of the $SrAl_2O_4:Eu^{2+}$ phosphor compared with the experimental data. The relative errors (in %) with respect to the experimental data are given in parentheses.

| | $SrAl_2O_4$:Eu1 | | $SrAl_2O_4$:Eu2 | |
|---|---|---|---|---|
| | Abs | Em | Abs | Em |
| Exp. data | 2.88 | 2.38 | 3.11 | 2.79 |
| DFT+U ($U_{eff}$ = 6 eV) | 3.23 (+12) | 2.56 (+8) | 3.46 (+11) | 2.78 (0) |
| DFT+U (Jia et al.)$^{\perp}$ | 2.97 (+3) | 2.32 (-3) | 3.17 (+2) | 2.55 (-9) |
| DFT+U (Hoang)$^{\#}$ | 4.00 (+39) | 3.38 (+42) | 4.23 (+36) | 3.52 (+26) |
| *HSE* (Hoang) | 3.69 (+28) | 3.05 (+28) | 3.91 (+26) | 3.42 (+23) |

$^{\perp}$ Reported values are calculated using DFT+U in the ABINIT package with U = 7.5 eV and J = 0.6 eV.[44]

$^{\#}$ Reported values are calculated using DFT+U in the VASP package with U = 7.5 eV and J = 0.6 eV.[21]

**Table 2.** The calculated absorption and emission energies (in eV) for the $Ba_4Si_6O_{16}:Eu^{2+}$ phosphor compared with the experimental data. The relative errors (in %) with respect to the experimental data are given in parentheses.

| | $Ba_4Si_6O_{16}$:Eu1 | | $Ba_4Si_6O_{16}$:Eu2 | |
|---|---|---|---|---|
| | Abs | Em | Abs | Em |
| Exp. data | 3.97 | 2.82 | 3.77 | 2.38 |
| DFT+U ($U_{eff}$ = 6 eV)$^{\perp}$ | 3.94 (-1) | 3.39 (+20) | 3.78 (0) | 3.02 (+27) |
| DFT+U ($U_{eff}$ = 6 eV)$^{\#}$ | 3.85 (-3) | 2.70 (-4) | 3.70 (-2) | 2.54 (+7) |

$^{\perp}$ The transitions are calculated between HOMO and LUMO+1 states.

$^{\#}$ The transitions correspond to average values between two transitions towards LUMO and LUMO+1 states.

**Table 2** shows similar results for the $Ba_4Si_6O_{16}:Eu^{2+}$ phosphor. Notably, the calculated absorption energies agree with experiment to better than 1%, while the emission energies deviate by more than 20%. To understand the source of these discrepancies, we must first clarify how the cDFT calculations are conducted. These calculations involve promoting an electron from the 4f(Eu) state to the 5d(Eu) state. While the 4f(Eu) bands are well localized, the 5d(Eu) bands are delocalized and hybridize with host states; consequently, the lowest empty Kohn-Sham state is not necessarily of dominant 5d(Eu) character. To quantify the 5d character of each candidate target state, we monitored the change in the number of 4f and 5d electrons within the Eu PAW sphere upon excitation. Because PAW projections capture only the electron density inside the sphere, they systematically underestimate the more delocalized 5d(Eu) population. Erreur ! Source du renvoi introuvable.**1** and Erreur ! Source du renvoi introuvable.**2** report these projections for $SrAl_2O_4:Eu^{2+}$ and $Ba_4Si_6O_{16}:Eu^{2+}$, respectively. For $SrAl_2O_4:Eu^{2+}$, excitation to the lowest empty state (LUMO) populates a band with substantial 5d(Eu) character, gaining ~0.3 electrons in the 5d projector while ~0.9 electrons are removed from the 4f projector. For $Ba_4Si_6O_{16}:Eu^{2+}$, the same predominantly-5d character is found instead at LUMO+1; excitation to the LUMO yields a smaller 5d gain of ~0.2 electrons. We report transitions up to LUMO+4 to illustrate the sensitivity of the result to the choice of unoccupied target state, though only LUMO and LUMO+1 are physically meaningful (**Table S3** and **Table S4**): higher target states give absorption energies well above the experimental value and cannot correspond to the lowest 5d(Eu) manifold. The associated optical transition energies are listed in **Table S3**. Notably, averaging the LUMO and LUMO+1 excitation energies (**Table 2**) brings the calculated energies within 7% of experiment. This underscores the limitations of the current cDFT approach, in which the photo-hole and photo-electron are pinned to specific bands without explicit treatment of the many-body nature of the excited state.

In summary, we propose the following mechanism for the PersL of $Ba_4Si_6O_{16}:Eu^{2+}$, $RE^{3+}$ (RE = Ho, Dy). Under UV illumination, $Eu^{2+}$ undergoes a $4f^7 \rightarrow 4f^6 5d^1$ optical transition, generating photoelectrons that delocalize into the *CB* and photoholes that are trapped at neutral oxygen vacancies $V_O^0$ associated with bridging $O_{in}$ sites along Si–O–Si corners, where the $V_O^0$ defect drives the formation of a short Si–Si bond across the vacancy. Through the negative-U ε(+2/0) transition, two holes are captured at the same vacancy, converting $V_O^0$ into $V_O^{2+}$ and breaking the Si–Si bond. The bridging-oxygen sites $V_{O3}$ and $V_{O4}$, with ε(+2/0) levels at 0.70 and 0.87 eV above the VBM, sit squarely in the trap-depth window required for room-temperature afterglow on timescales of seconds to minutes, and constitute the active hole traps in the PersL mechanism. The photoelectrons released into the *CB* are predominantly captured by the $RE^{3+}$ co-dopants, forming $RE^{2+}$ ($Ho^{2+}$ or $Dy^{2+}$; such metastable $RE^{2+}$ species were experimentally evidenced in $RE^{3+}$-doped $SrAl_2O_4:Eu^{2+}$, see ref. [53]); however, the corresponding ε(+1/0) levels lie 1.22 eV (Ho) and 1.04 eV (Dy) below the *CBM*, outside the room-temperature PersL window, so $RE^{2+}$ functions as an auxiliary deep reservoir that increases the host storage capacity rather than as the kinetics-controlling electron trap. PersL then arises from electron-hole recombination at the Eu site following thermal release of the holes from the $V_O^{2++}$ traps; this hole-release step is the rate-limiting process at room temperature, with the deeper $RE^{2+}$ reservoir feeding the recombination on much longer timescales and accounting for the enhanced and prolonged afterglow observed under $RE^{3+}$ co-doping.

### 3.4 Quantitative and mechanistic description of the roles of stress nature and magnitude

Having identified the active point defects and charge carriers in the PersL mechanism of $Ba_4Si_6O_{16}$:$Eu^{2+}$, $RE^{3+}$ (RE = Ho, Dy), it is essential to quantify the impact of the applied mechanical stress on the associated charge-transition levels. As a first step, we consider hydrostatic stresses, both compressive (negative) and tensile (positive), ranging from -6 to +4 GPa. Although larger than those reported in previous studies, this range is chosen to make the resulting changes in trap depth clearly visible and to represent the substantial local stresses that may develop at the interface between the active crystals and the surrounding matrix (glass or epoxy resin) during testing. The results are shown in **Figure 7(a)**, with the unstressed supercell (0 GPa) included as a reference.

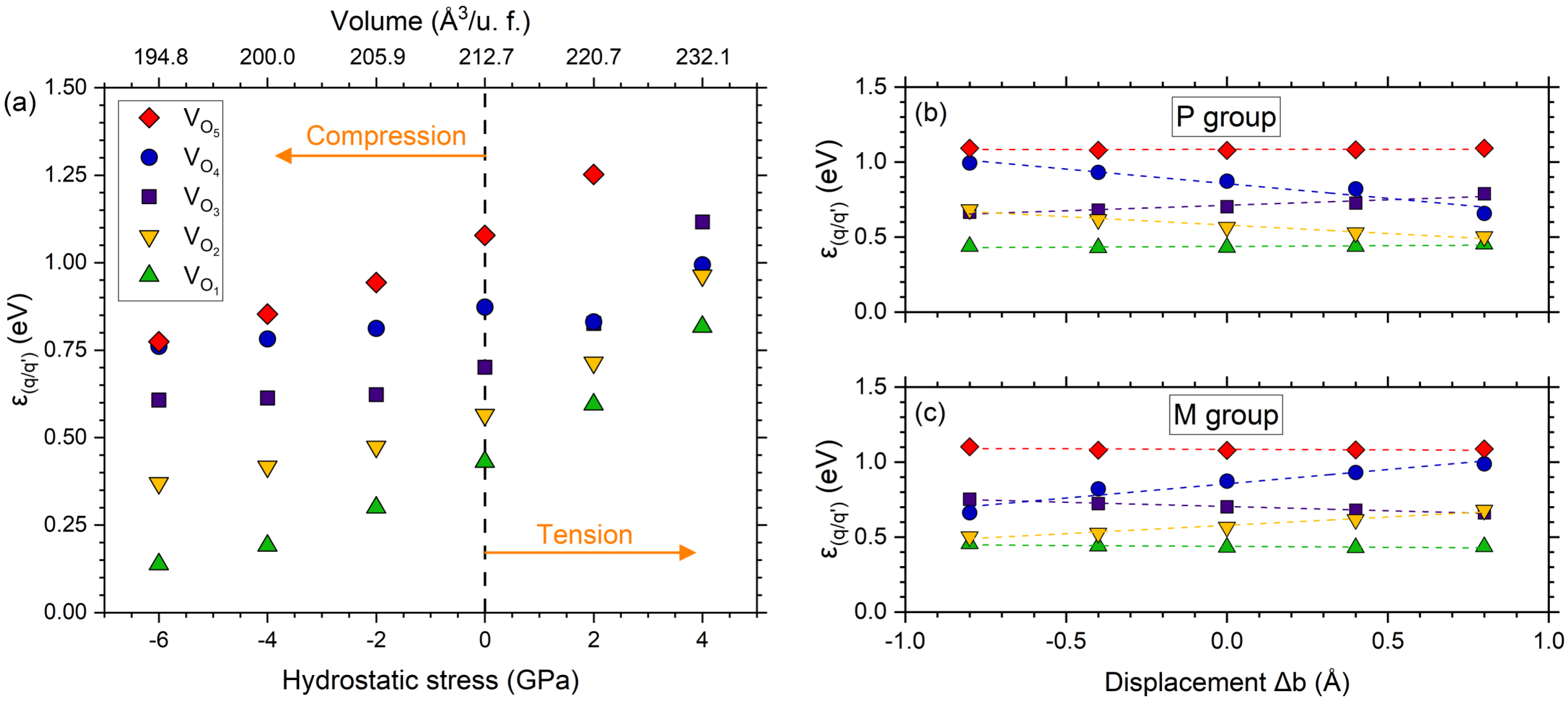

**Figure 7.** Evolution of the thermodynamic charge-transition levels of oxygen vacancies in $Ba_4Si_6O_{16}$ under mechanical perturbations. The reported levels correspond to ε(2+/0) for the inner-oxygen vacancies $V_{O1}$, $V_{O2}$, $V_{O3}$, and $V_{O4}$, and to ε(2+/1+) for the outer-oxygen vacancy $V_{O5}$. (a) Transition levels as a function of applied hydrostatic stress, ranging from −6 GPa (compression) to +4 GPa (tension); the corresponding equilibrium unit-cell volumes are indicated on the upper axis, and the dashed vertical line marks the unstrained structure (0 GPa, 212.7 $Å^3$ per formula unit). For $V_{O5}$, ε(2+/1+) reaches 2.11 eV under 4 GPa of hydrostatic tension and falls outside the plotted range. (b, c) Transition levels as a function of an anisotropic displacement Δb applied along the *b* axis, shown separately for the two symmetry-distinct subgroups of inequivalent oxygen sites: the P group (b) and the M group (c). Symbols and colors: $V_{O1}$ (green triangles), $V_{O2}$ (yellow inverted triangles), $V_{O3}$ (purple squares), $V_{O4}$ (blue circles), and $V_{O5}$ (red diamonds); dashed lines in panels (b) and (c) are guides to the eye.

The overall crystal structure and the local reorganizations around the point defects are largely unaffected by the applied hydrostatic stress, and the formation energies vary by only a few tenths of an eV at most. The trap depths, however, shift more systematically: each energy level deepens almost linearly under tensile stress and shallows almost linearly under compressive stress, with a total variation of ~0.1 eV across the −6 to +4 GPa range. This trend agrees with a previously reported phenomenological model that assumes (i) a linear decrease in trap depth with increasing compressive stress, and (ii) a reduction in trap depth of ~0.07 eV to account for the observed mechanoluminescence response. It is also consistent with the mechanoluminescence behavior previously reported under isostatic pressure: when $Ba_4Si_6O_{16}:Eu^{2+}, Ho^{3+}$ is subjected to compressive loading, the trap depths decrease and the corresponding thermal-release rate increases, which accounts for the enhanced luminescence intensity during loading. In the case of a three-point bending test, the situation is more complex because shear occurs equally on both sides of the neutral axis, while tensile stresses develop on the lower side and compressive stresses on the upper side. The way in which the deviatoric and hydrostatic components of the stress combine to modify the depth of the trap requires further investigation. The coordinating (or near neighboring) ions (around a peculiar point defect site) may get closer in one direction and move away in another one. This causes subtle changes of the crystal field strength, as discussed in ref. [54].

As a second step, we consider the role of shear stresses, *i.e.*, the deviatoric contribution. Among the six possible shear displacements, four lower the space group from $P2_1/c$ to $P\bar{1}$, while the remaining two, gliding the $(a, b)$ plane along the $a$-axis or the $(b, c)$ plane along the $c$-axis (*i.e.*, the $\gamma_{13}$ distortion), leave the space group unchanged. All six preserve centrosymmetry, so shear stress alone cannot induce piezoelectricity in this phase. The most energetically favorable of the six is the gliding of the $(a, b)$ plane gliding along $b$ ($\gamma_{23}$ distortion) (see Erreur ! Source du renvoi introuvable.**(a-b)**), which is parallel to the silicate chains extending along that axis. We therefore applied displacements $\Delta b$ in the range -0.8 to +0.8 Å to this plane. The shear modulus of pure $Ba_4Si_6O_{16}$ has not been reported, but we measured a value of 33 ± 2 GPa for the $Ba_4Si_6O_{16}:Eu^{2+}, Ho^{3+}$-containing glass-ceramic. Assuming a comparable value of ~30 ± 5 GPa for the parent crystal, $\Delta b = +0.8$ Å corresponds to a shear stress of 1.7 ± 0.3 GPa. Gliding the $(a, b)$ plane along b breaks the $2_1$ screw axis, so each crystallographically inequivalent oxygen site (multiplicity 4) splits into two inequivalent positions (multiplicity 2). We label the resulting two sets of 8 sites the “P group” and the “M group”. The effect of this displacement on the charge-transition levels of the two groups is shown in **Figure 7(b,c)** for the oxygen vacancies identified above. Within the M group, increasing $\Delta b$ decreases the trap depth only for $V_{O_3}$, leaves $V_{O_1}$ and $V_{O_5}$ unchanged, and increases the trap depth for $V_{O_2}$ and $V_{O_4}$. Additionally, the M and P groups display opposite trends. This stands in marked contrast with the hydrostatic case: under shear, only a fraction of the energy levels becomes shallower, and to a much smaller extent. This behavior is consistent experimental measurements showing that the trap depths, and hence the luminescence intensity, remain largely unaffected, or only weakly affected, by shear stress. Two experimental observations follow directly (i) no mechanoluminescence intensity is detected during torsion tests, and (ii) the mechanoluminescence intensity recorded during uniaxial compression tests is roughly two orders of magnitude weaker than under isostatic pressure.

This behavior likely stems from the crystal structure of $Ba_4Si_6O_{16}$ and from the nature of the trapping defects involved. $Ba_4Si_6O_{16}$ has a quasi-1$D$ topology, in which silicate ribbons are separated by $BaO_8$ sheets. As a result, the local environment of an oxygen vacancy is only weakly perturbated when the silicate chains, *i.e.*, the (a, b) plane, glide along b. More generally, a crystal should exhibit mechanoluminescence during torsion only if both the dimensionality of its host structure and the nature of its trapping defects allow shear stress to alter the trap environment. Direct support for this hypothesis remains limited in the literature, since mechanoluminescence is most commonly characterized by diametral compression tests, which involves a complex stress state, for practical reasons; nevertheless the hypothesis is consistent with the observation that 3*D*-structured phosphors such as $SrAl_2O_4:Eu^{2+}$, $Dy^{3+}$ [55] and $ZnS:Cu^{+}$ [56] exhibit mechanoluminescence under torsion, whereas the quasi-1*D* $Ba_4Si_6O_{16}:Eu^{2+}$, $Ho^{3+}$ does not. In a 3*D* crystal structure, shear stress modifies the local environment of every site, shifting all trap depths and producing the corresponding luminescence response; in a 1D framework the chains glide as relatively rigid units, leaving most trap environments are largely intact.

Various mechanisms have been proposed to account for mechanoluminescence [1 4]. (i) The mechanical energy supplied during loading could in principle promote trapped charge carriers across their trap-depth barrier directly. However, Feng *et. al* [1] estimated that, in a typical mechanoluminescence experiment, each atom receives only ~$4{\cdot}10^{-6}$ eV, four order of magnitude smaller than the thermal energy $k_B T$ = $26{\cdot}10^{-3}$ eV at room temperature, and therefore far too small to account for the observed luminescence response. (ii) Piezoelectricity can lower the effective trap depth and thereby accelerate carrier release. In $SrAl_2O_4$, for example, a shear stress of ~4 GPa is calculated to induce a substantial electric polarization (~1 $\mu C{\cdot}cm^{-2}$, assuming a shear modulus of 41 GPa [57]), which bends the valence and conduction bands and brings the trap states closer to the band edges. This picture also applies to nominally centrosymmetric crystals, where local symmetry breaking around defects can produce a local piezoelectric response even when the bulk space group forbids one[1].

By contrast, our calculations show unambiguously that defect-containing $Ba_4Si_6O_{16}$ supercells develop no piezoelectric field under either hydrostatic or shear stress. The observed changes in the charge-transition levels must therefore arise solely from the structural reorganization around the point defects under stress, as illustrated in **Figure 8(a-b)**. Indeed, a shorter Si(1)-Si(2) distance systematically corresponds to a shallower trap, and conversely. Plotting the trap depth directly against the Si(1)-Si(2) distance directly reveals how each ${V_{O_x}}^0$ defect responds to a given stress mode and identifies which levels are most sensitive to which type of loading. Two regimes emerge. Under shear, several levels respond only weakly (${V_{O_3}}^0$) or not at all (${V_{O_1}}^0$). Under hydrostatic stress, by contrast, every Si(1)–Si(2) distance varies substantially and every trap depth decreases, the ${V_{O_1}}^0$ and ${V_{O_2}}^0$ defects most of all, and the magnitude of the depth shift is markedly larger than under shear at the same Si(1)–Si(2) distance. A further trend visible in **Figure 8(c)** is that levels closer to the *VBM* correspond to larger *ICOBI* values, *i.e.*, to stronger Si–Si bonding across the reconstructed vacancy site.

Two distinct microscopic contributions therefore govern stress-induced changes in trap depth in mechanoluminescent crystals: (i) the generation of a piezoelectric field, as in $SrAl_2O_4:Eu^{2+}$, $Dy^{3+}$, and (ii) structural reorganization around the carrier-trapping point defects, as in $Ba_4Si_6O_{16}:Eu^{2+}$, $Ho^{3+}$. The mechanoluminescence intensity scales with the dominant contribution: a stronger induced piezoelectric field, or a higher local stress sensitivity of the trap environment, produces brighter emission. A complete understanding of the mechanoluminescence mechanism in any given material therefore requires both contributions to be considered together.

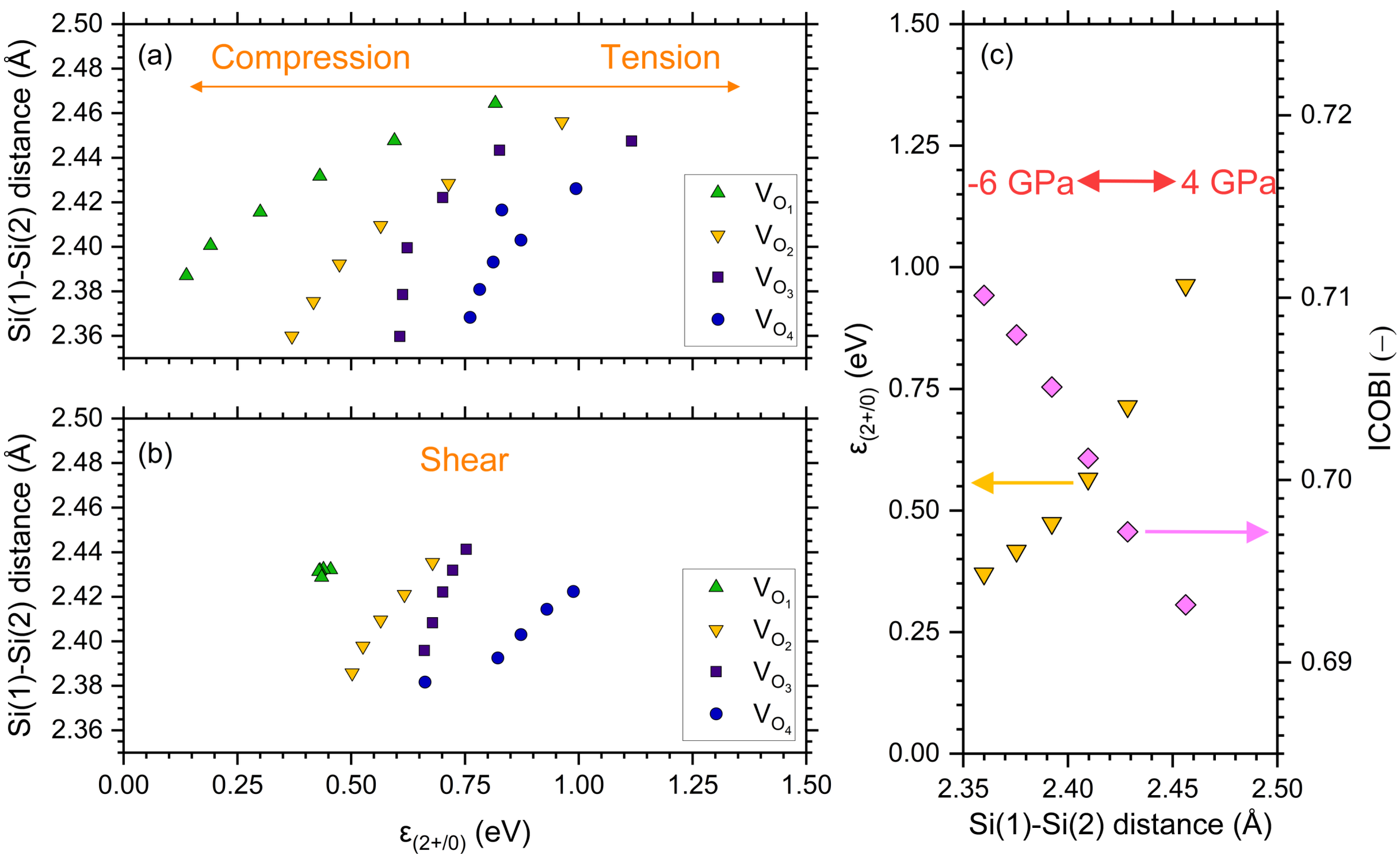

**Figure 8.** Correlation between the thermodynamic charge-transition level ε(2+/0) of neutral oxygen vacancies ($\boldsymbol{V_{O_x}^{0}}$) and the Si(1)–Si(2) distance across the Si(1)–$O_x$–Si(2) bridge from which the oxygen atom $O_x$ has been removed. (a) Data obtained under applied hydrostatic stress, ranging from −6 GPa (compression) to +4 GPa (tension); each point corresponds to a single (defect, stress) configuration. (b) Data obtained under an anisotropic shear-like displacement Δb applied to the M-group oxygen sites. In both panels, the four inner-oxygen vacancies $V_{O1}$, $V_{O2}$, $V_{O3}$, and $V_{O4}$ follow a common trend in which ε(2+/0) increases monotonically with the Si(1)–Si(2) distance, indicating that the transition level is governed primarily by the geometry of the Si–Si reconstruction across the vacancy rather than by the identity of the oxygen site. (c) Correlation between ε(2+/0) of the $\boldsymbol{V_{O_2}^{0}}$ defect (yellow inverted triangles, left axis) and the integrated crystal orbital bond index (*ICOBI*) of the corresponding reconstructed Si(1)–Si(2) bond (pink diamonds, right axis), as functions of the Si(1)–Si(2) distance, for hydrostatic stresses spanning −6 to +4 GPa. The opposite trends of ε(2+/0) and the *ICOBI* confirm that shorter Si–Si distances correspond to stronger Si–Si bonding and to deeper transition levels. Symbols: $V_{O1}$ (green triangles), $V_{O2}$ (yellow inverted triangles), $V_{O3}$ (purple squares), $V_{O4}$ (blue circles).

## 4. Conclusions

In this study, the formation energies and charge-transition levels of the principal native and substitutional point defects in $Ba_4Si_6O_{16}$ were computed using first-principles methods. This analysis identifies the most probable trapping-detrapping processes underlying persistent luminescence in crystalline $Ba_4Si_6O_{16}$:$Eu^{2+}$, $Ho^{3+}$ and in the corresponding glass-ceramic systems, namely, holes trapping at bridging-oxygen vacancies, with trap depths matching those extracted from thermally stimulated luminescence experiments. The same hole-trap-on-$V_O$ motif may operate in other silicate phosphors built on corner-sharing $SiO_4$ networks.

Hydrostatic and shear loading produced strikingly different responses. Under hydrostatic stress, every trap depth decreases, in agreement with the experimental observation of mechanoluminescence under isostatic pressure and with established phenomenological models, defining hydrostatic compression as the dominant driving force for mechanoluminescence in this system. Under shear, by contrast, the trap-depth shifts are smaller and of mixed sign, a direct consequence of the quasi-1D silicate-chain topology of $Ba_4Si_6O_{16}$, and consistent with the experimentally observed absence of mechanoluminescence under torsion. Structural dimensionality therefore plays a critical role in dictating both the magnitude and the directionality of the mechanoluminescence response.

Two distinct microscopic contributions to mechanoluminescence emerge from these results: a piezoelectric contribution, restricted to non-centrosymmetric hosts, and a contribution from local structural reorganization around trapping defects under stress, which is operative in any host. Together, they provide a more complete picture of stress-stimulated charge-carrier dynamics, with implications for both crystalline and amorphous mechanoluminescent materials. The implication for materials design is direct: bright mechanoluminescent systems are obtained by combining a host structure that supports piezoelectric coupling with point defects whose trap environment is highly sensitive to the relevant stress mode, while quasi-1D hosts containing shear-insensitive defects will exhibit a strongly anisotropic response.

**Acknowledgements**

We acknowledge financial support from Région Bretagne.

**Author declarations**

**Conflicts of interests**
The authors have no conflicts to disclose.

**Author Contributions**
**Alexis Duval:** Conceptualization (lead); Data curation (lead); Investigation (lead); Methodology (lead); Writing – original draft (lead). **Xavier Rocquefelte:** Conceptualization (lead); Data curation (lead); Investigation (lead); Methodology (lead); Writing – original draft (lead). **Yann Gueguen:** Investigation (equal); Writing – review & editing (equal). **Patrick Houizot:** Investigation (equal); Supervision (equal); Writing – review & editing (equal). **Tanguy Rouxel:** Investigation (equal); Supervision (equal); Writing – review & editing (equal).

**Data availability**

The data that support the findings of this study are available within the article and from the corresponding author upon reasonable request.

**ORCID**

Alexis Duval iD https://orcid.org/0000-0001-9192-2454

Xavier Rocquefelte iD https://orcid.org/0000-0003-0191-2354

Yann Gueguen iD https://orcid.org/0000-0001-6348-3060

Patrick Houizot iD https://orcid.org/0000-0002-6360-6996

Tanguy Rouxel iD https://orcid.org/0000-0002-9961-2458